\def\figsize{9.5cm}

\def\rn{}
\def\nn#1 #2{#2. #1}                            % Name with 1 initial
\def\nnn#1 #2 #3{#2. #3. #1}                    % Name with 2 initials
\def\nnnn#1 #2 #3 #4{#2. #3. #4 #1}             % Name with 3 initials
\def\nnnnn#1 #2 #3 #4 #5{#2. #3. #4 #5. #1}     % Name with 4 initials
\def\rf#1;#2;#3;#4;#5 {{\frenchspacing\par\rn#1, #3 {\bf #4}, #5 (#2). \par}}
\def\rrf#1;#2;#3;#4;#5 {{\frenchspacing\rn#1, #3 {\bf #4}, #5 (#2);~}}
\def\rrrf#1;#2;#3;#4;#5 {{\frenchspacing\rn#1, #3 {\bf #4}, #5 (#2).}}
\def\rg#1;#2;#3;#4;#5;#6 {{\frenchspacing\par\rn#1, #3 {\bf #4}, #5 (#2). \par}}
\def\rfbook#1;#2;#3;#4;#5 {{\frenchspacing\par\rn#1, {\it #3} (#5, #4, #2).\par}}
\def\rfprep#1;#2;#3 {{\par\frenchspacing\rn#1, #3 (#2).\par}}
\def\rrfprep#1;#2;#3 {{\frenchspacing\rn#1, #3 (#2);~}}
\def\rrrfprep#1;#2;#3 {{\frenchspacing\rn#1, #3 (#2).}}
\def\rfproc#1;#2;#3;#4;#5;#6 {{\frenchspacing\par\rn#1 #2, in {\it #3}, ed. #4 (#5: #6)\par}}
\def\rfprocp#1;#2;#3;#4;#5;#6;#7 {{\frenchspacing\par\rn#1 #2, in {\it #3}, ed. #4 (#5: #6), p#7\par}}
\def\rg#1;#2;#3;#4;#5;#6 {\par\rn#1 #2, {\it #3}, {\bf #4}, #5 (``#6'') \par}
\def\rf#1;#2;#3;#4;#5 {\par\rn#1, {\it #3}, {\bf #4}, #5 (#2)\par}
\def\rfbook#1;#2;#3;#4;#5 {{\frenchspacing\par\rn#1, {\it #3} (#4: #5, #2)\par}}
\def\rfproc#1;#2;#3;#4;#5;#6 {{\frenchspacing\par\rn#1 #2, in {\it #3}, ed. #4 (#5: #6)\par}}
\def\rfprocp#1;#2;#3;#4;#5;#6;#7 {{\frenchspacing\par\rn#1 #2, in {\it #3}, ed. #4 (#5: #6), p#7\par}}
\def\rfprep#1;#2;#3  {{\par\rn#1, #3, #2\par}}
\def\rfprepp#1;#2;#3 {{\par\rn#1 #2, #3\par}}

\def\beq#1{\begin{equation}\label{#1}}
\def\eeq{\end{equation}}
\def\beqa#1{\begin{eqnarray}\label{#1}}
\def\eeqa{\end{eqnarray}}

\def\spose#1{\hbox to 0pt{#1\hss}}
\def\simlt{\mathrel{\spose{\lower 3pt\hbox{$\mathchar"218$}}
     \raise 2.0pt\hbox{$\mathchar"13C$}}}
\def\simgt{\mathrel{\spose{\lower 3pt\hbox{$\mathchar"218$}}
     \raise 2.0pt\hbox{$\mathchar"13E$}}}
\def\simpropto{\mathrel{\spose{\lower 3pt\hbox{$\mathchar"218$}}
     \raise 2.0pt\hbox{$\propto$}}}

\def\ed{\end{document}}

\def\beq#1{\begin{equation}\label{#1}}
\def\eeq{\end{equation}}
\def\beqa#1{\begin{eqnarray}\label{#1}}
\def\eeqa{\end{eqnarray}}

\documentclass[twocolumn,amsmath,nofootinbib]{revtex4} % For astro-ph
\usepackage{graphicx}% Include figure files
\usepackage{dcolumn}% Align table columns on decimal point
\usepackage{bm}% bold math
\usepackage{url}

\def\subfigureA#1{
\leavevmode
\hbox{#1}
}

\newcommand{\m}{\medbreak}

\newcommand{\no}{\noindent}
\newcommand{\EQ}{\begin{equation}}
\newcommand{\EQA}{\begin{eqnarray}}
\newcommand{\eqa}{\end{eqnarray}}

\newcommand{\AR}{\renewcommand {\arraystretch}{1.5}
\begin{array}{l}}
\newcommand{\bAR}{\renewcommand {\arraystretch}{2}
\begin{array}{l}}
\newcommand{\ARc}{\renewcommand {\arraystretch}{1.5}
\begin{array}{c}}
\newcommand{\bARc}{\renewcommand {\arraystretch}{2}
\begin{array}{c}}
\newcommand{\ar}{\end{array} \renewcommand {\arraystretch}{1}}

\begin{document}

%%%%%%%%%%%%%%%%%%%%%%%%%%%%%%%%%%%%%%%%%%%%%%%%%%%%%%%%%%%%%
%%%%%%%%%%%%%% new (documentclass) %%%%%%%%%%%%%%%%%
\input{epsf.sty}
\begin{titlepage}
\vspace*{.15in}
%\noindent
%PHYSICAL REVIEW LETTERS
%\vskip .20in
%\noindent

%%%%%
%%%%%
%\today
%\vspace{0.2in}
%\begin{flushright}
%%draft1 AE+CT\\
%\end{flushright}
\vspace*{0.5cm}
\begin{center}
{\Large \bf Probing Dark Energy with Supernovae : \\ \m
a concordant or a convergent model?}\\

\m
\vspace*{1.2cm}
{\bf J.-M. Virey$^1$, A. Ealet$^2$, C. Tao$^2$, A. Tilquin$^2$, A. Bonissent$^2$, D. Fouchez$^2$ and P. Taxil$^1$ }

\vspace*{1.2cm}
$^1$Centre de Physique Theorique$^*$, CNRS-Luminy,
Case 907, F-13288 Marseille Cedex 9, France \\
and Universite de Provence \\ 
\vspace*{0.5cm}
$^2$Centre de Physique des Particules de Marseille$^+$, CNRS-Luminy,
Case 907, F-13288 Marseille Cedex 9, France

\vspace*{2.5cm}
{\bf Abstract} \\
\end{center}

We present a revised interpretation of recent
analysis of supernovae data. 
We evaluate the effect of the priors on the extraction
of the dark energy equation of state.
We find that the conclusions depend 
strongly on the $\Omega_M$ prior value and on its uncertainty,
and show that a biased fitting procedure 
applied on non concordant simulated data
can converge to the "concordance model". 
Relaxing the prior on $\Omega_M$ points to other sets of solutions, 
which are not excluded by observational data.

\vfill
\begin{flushleft}
\m\no
Number of figures : 6\\

\m\no
July 2004\\
CPT-2004/P.034\\
CPPM-P-2004-02\\
\m\no
anonymous ftp or gopher : cpt.univ-mrs.fr\\ \no
E-mail : virey@cpt.univ-mrs.fr\\ \m

\no ------------------------------------\\ \m
\no $^*$``Centre de Physique Th\'eorique'' is UMR 6207 - ``Unit\'e Mixte
de Recherche'' of CNRS and of the Universities ``de Provence'',
``de la M\'editerran\'ee'' and ``du Sud Toulon-Var''- Laboratory
affiliated to FRUMAM (FR 2291).\\ \m

\no $^+$``Centre de Physique des Particules de Marseille'' is UMR 6550 
of CNRS/IN2P3 and of the University
``de la M\'editerran\'ee''.
%$^{\ast}$Unit\'e  de Recherche yyyy
% \\
\end{flushleft}
\newpage
\end{titlepage}

%%%%%%%%%%%%%%%%%%%%%%%% fin old (documentstyle) %%%%%%%%%%%%%
%%%%%%%%%%%%%%%%%%%%%%%%%%%%%%%%%%%%%%%%%%%%%%%%%%%%%%%%%%%%%%%%%%%%%%%%%%%%
                   %%%%%%%%%%%%%%%%%%%%%%%%%%%%%%%%%%
   
%\pagestyle{myheadings}
%\markright{draft2-AE+CT}  
%%%%%%%%%%%%%%%%%%%%%%%%%%%%%%%%%%%%                         

%\section{Introduction}

\indent
%\m

The existence and nature of dark energy is one of the most challenging issues 
of physics today.
The publication of high redshift supernovae discovered by the Hubble Space
Telescope, by the SCP collaboration \cite{newSCP}
and recently by Riess et al. \cite{Riess04}, 
has been interpreted as agreement of the data with the 
so named $\Lambda CDM$  "concordance model" 
($\Omega_M \approx 0.3$, $\Omega_{\Lambda} \approx 0.7$, $w=p/{\rho}=-1$).
We have reconsidered some conclusions in the light of our previous analysis of simulated
data \cite{biasSN}. 

Riess et al.\cite{Riess04} 
have selected 157 well measured SNIa, which they
call the "gold" sample, a set of data we will use throughout this paper. 
Assuming a flat Universe ($\Omega_T = 1$) they conclude that:
{\it i)} 
Using the strong prior $\Omega_M = 0.27 \pm
0.04$, a fit to a static dark energy equation of state yields    
-1.46$<w<$-0.78 ($95\% CL$);
{\it ii)} Looking at a possible redshift dependence of $w(z)$ 
(using $w(z)= w_0 + w_1 z$), the data with the strong prior 
%$\Omega_M = 0.27 \pm 0.04$
indicate that the region $w_1<0$ and especially the quadrant ($w_0 > -1$ and $w_1 < 0$) 
are the least
favoured. They reject large time variation and are compatible with the concordance model. \\
%They point out a gain factor of more than 8 on the accuracy of the measurement.\\
 
We have shown in \cite{biasSN} that it is unavoidable to get 
some ambiguities when trying to fit a particular fiducial cosmology with a "wrong" model.
This "bias problem" has been mentioned several times in the literature, 
see {\it e.g.}\cite{Maor,WA,Gerke,LinderB}.
In this letter, we explore the effect of the $\Omega_M$ prior on the determination of $w(z)$.

 Following \cite{biasSN}, we assume a flat universe and keep the same parametrisation
of $w(z)$ as in \cite{Riess04}, for the sake of comparison.
We call 3-fit (4-fit) the fitting procedure
which involves the 3 (4) parameters $M_S$, $\Omega_M$ and $w_0$ ($M_S$, $\Omega_M$, $w_0$ and $w_1$),
$M_S$ being a normalisation
parameter (see \cite{biasSN} for definitions and formulae).
We have performed 3-fits
and 4-fits and compared the results in different cases, varying the central
value and the uncertainty on the $\Omega_M$ prior.
\begin{table}[h]
\vspace*{-0.4cm}
\caption{\protect\footnotesize Fit results obtained using the gold data from \cite{Riess04} for various 
fitting procedures. The $\chi^2$ is very stable, it is around 173 (for 157 SNIa)
for all procedures except
for the 3-fit with the strong prior $\Omega_M =0.27\pm 0.04$ where $\chi^2\approx 176$.}
\m
%\vspace*{0.2cm}
\label{tab:riess}
\begin{tabular}{ |c|c|c|c|c|}
\hline 
 Fit &  $\Omega_M$ prior  &   $\Omega_M$
& $w_0$  &  $w_1$   \\
\hline \hline
3-fit & no   & $0.48 \pm 0.06$  & $ -2.2 \pm 0.95$  & /   \\
3-fit &  $0.27 \pm 0.2$  & $0.45 \pm 0.07$  & $-1.9 \pm 0.73$  & /  \\
3-fit &  $0.50 \pm 0.2$ & $0.48 \pm  0.06$  & $-2.3 \pm 0.94$ & /   \\
3-fit & $0.27 \pm 0.04$ & $0.28 \pm 0.04$  & $-1.0 \pm 0.15$  & /    \\
3-fit & $0.50 \pm 0.04$ & $0.49 \pm 0.03$ & $ -2.5 \pm 0.77$  & / \\
\hline \hline
4-fit & no & $0.48 \pm 0.20$ & $ -2.2 \pm 1.34$  & $0.12 \pm 23 $  \\
4-fit & $0.27 \pm 0.2$ & $0.35 \pm 0.18$ & $- 1.6\pm 0.80$  & $1.74 \pm 1.3$  \\
4-fit & $0.50 \pm 0.2$ & $0.49 \pm 0.20$ & $-2.6 \pm 1.20$  & $1.60 \pm 18$    \\
4-fit & $0.27 \pm 0.04$ & $0.28 \pm 0.04$  & $-1.3 \pm 0.26$  & $1.50 \pm 0.84$   \\
4-fit & $0.5 \pm 0.04$ &  $0.49 \pm 0.04$  & $-2.6 \pm 1.40 $ & $ 0.95 \pm 10 $     \\
\hline
\end{tabular}
%\end{center}
%\vspace*{-0.4cm}
\end{table}

Applying no prior or the strong prior on $\Omega_M$ (lines 1, 4 and 9 of the Table),
we recover the results obtained by Riess et al.\cite{Riess04}.
Nevertheless, some interesting points can be underlined:

\no $\bullet$ With no prior or a weak prior on $\Omega_M$, the preferred $\Omega_M$ values 
are always greater than 0.3.

\no $\bullet$ Without any assumption on $\Omega_M$ nor $w_1$, the error on $\Omega_M$ is close to 0.2
(line 6 of Table I).

\no $\bullet$ Changing the central value of the $\Omega_M$ prior leads to a change in the $w_0$ values
of more than 1$\sigma$. The $w_0$ values are strongly correlated to $\Omega_M$ and are thus
 always smaller than the $\Lambda {CDM}$ value, when the strong prior on $\Omega_M$ is relaxed. 
$\chi^2$ is very stable but the correlation matrix can vary a lot for the 4-fits 
and the ($w_0$,$w_1$) solution.

\no $\bullet$ If the $\Omega_M$ prior is strong, the conclusion on $w_0$ depends on the prior value : 
for $\Omega_M$=0.27,  $w_0$ is forced to values compatible with -1, in particular for the 3-fit 
 and the errors are strongly reduced. For $\Omega_M$=0.5,  
$w_0$  is more negative and the errors are significantly larger.

\no $\bullet$ The only cases where ``reasonable'' errors can be found on $w_1$ 
% (ie, the error on $w_1$ < is less than $w_1$) 
occur for $\Omega_M$ around 0.3.

To illustrate these points,
Figure~\ref{fig:omdata} shows the results in
the ($\Omega_M$,$w_0$) plane for the 3-fits(left) and the 4-fits(right), 
using no prior on $\Omega_M $ or two strong priors with the two central values:
0.27 and 0.5. As expected the contours strongly depend on the procedure used to analyse
the data. 
For instance, the $95\% CL$ contours for the two strong prior cases are disconnected.
However,
we note that $\Omega_M < 0.6$ is valid for all procedures, hence it is one of the strong
conclusions from present SN data. \\
\def\figsize{4.5cm}
\begin{figure}[ht]
\vspace{-.7truecm}
\hspace{-.6truecm}
\begin{tabular}[t]{c c}
\centerline{\subfigureA{\epsfxsize=\figsize\epsffile{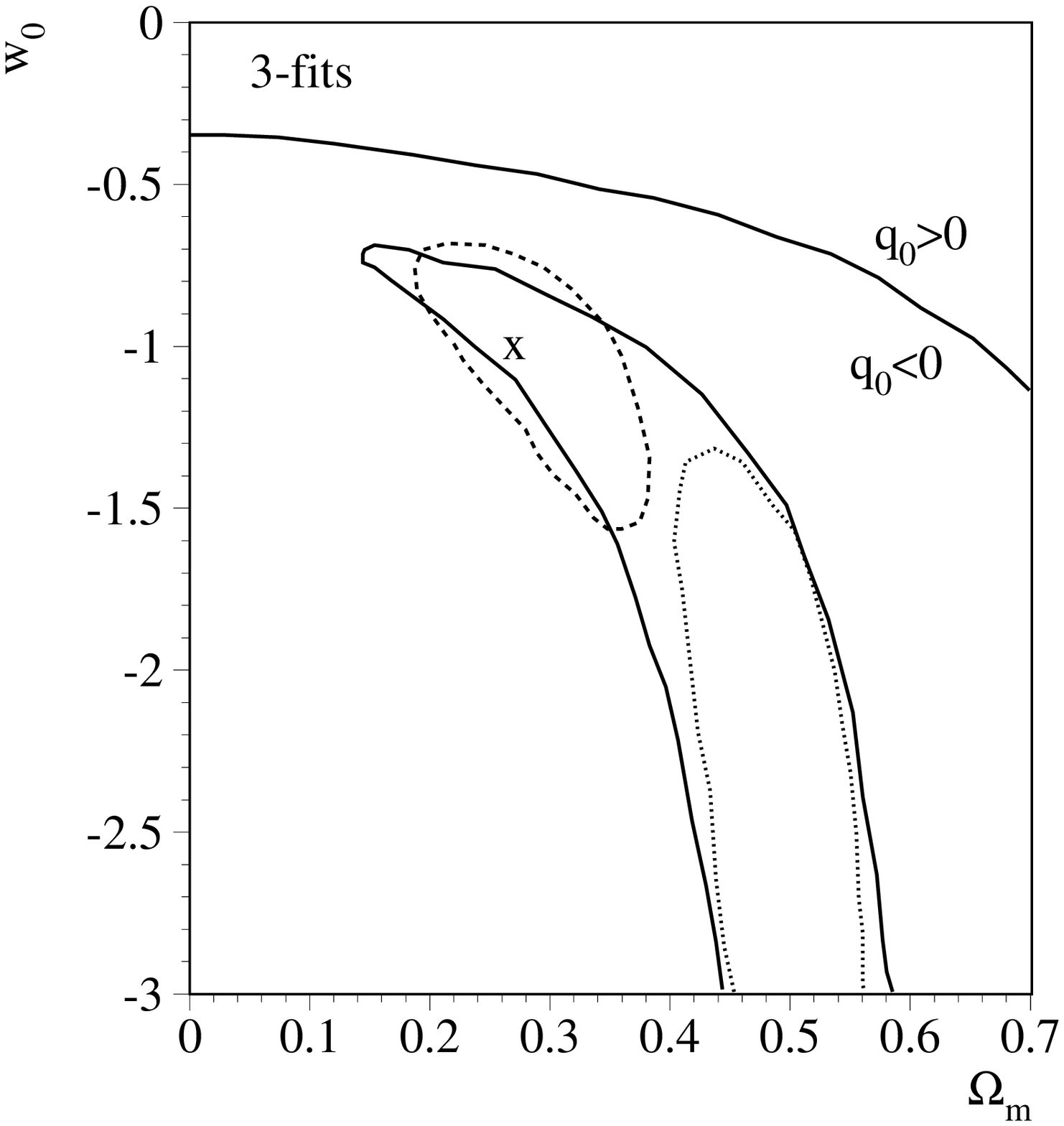}}
\subfigureA{\epsfxsize=\figsize\epsffile{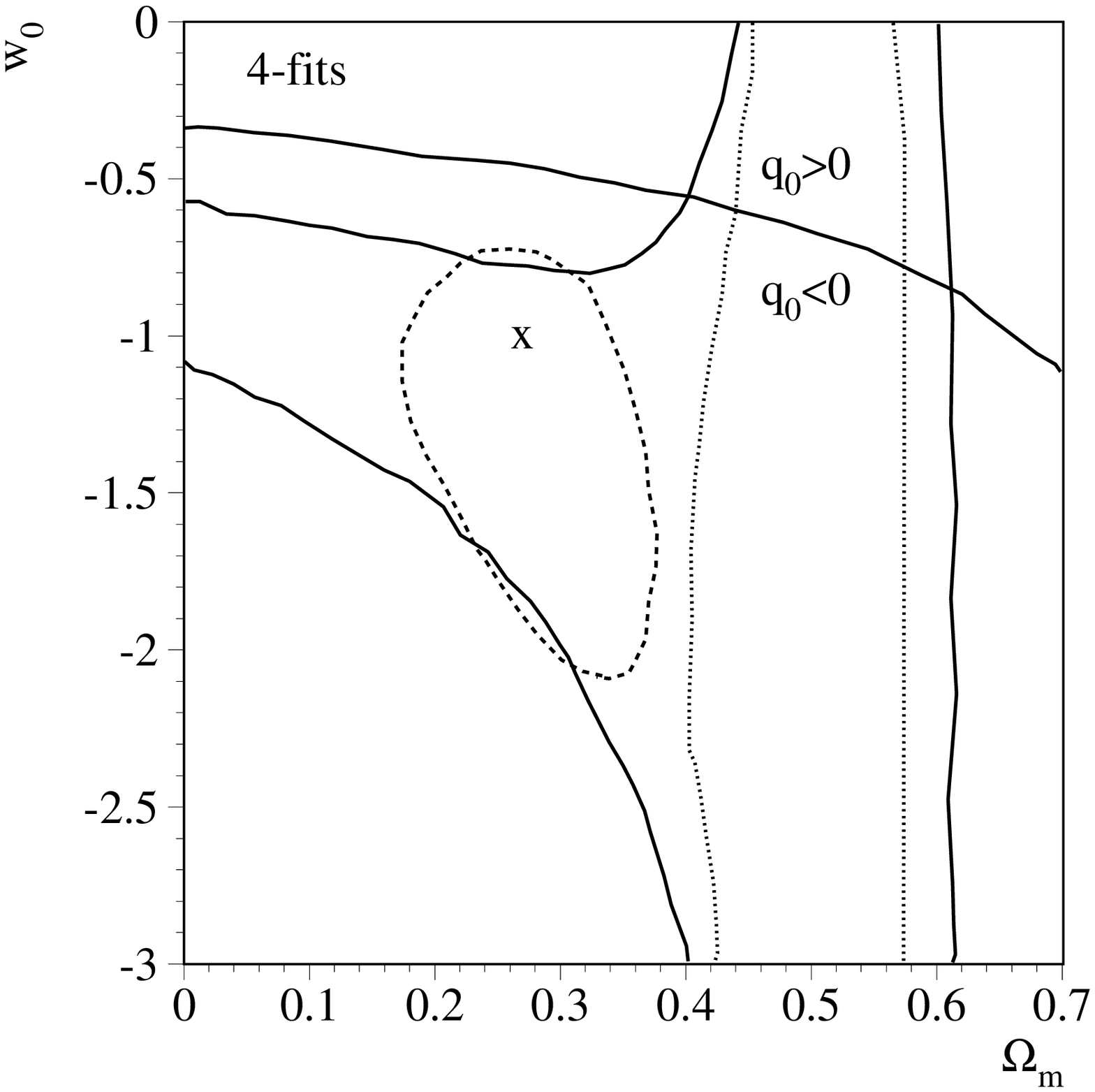}}}
\end{tabular} 
\vspace{-1.3truecm}
\caption{\footnotesize $95\% CL$ contours for 3-fits(left) and 4-fits(right) with
no prior on $\Omega_M$ (plain) and two strong priors $\Omega_M =0.27(0.5) \pm 0.04$ (dashed (dotted)). The (x) indicates
the $\Lambda CDM$ point ($\Omega_M=0.27, w_0=-1, w_1=0$). 
The plain line separates accelerating ($q_0<0$) from decelerating ($q_0>0$) models.
}
    \label{fig:omdata} 
%    \label{omdata} 
\vspace{-.5truecm}
\end{figure}
\def\figsize{9.5cm}
%%%%%%%%%%%%%%%%%%%%%%%%%%%%%%%%%%%%%%%%%%%%%%%%%%%%%%%%%%%%%%%%%%%%%%%%%%%%%%%%%%%%%%%%%%%%%%%%%%
%%%%%%%%%%%%%%%%%%%%%%%%%%%%%%%%%%%%%%%%%%%%%%%%%%%%%%%%%%%%%%%%%%%%%%%%%%%%%%%%%%%%%%%%%%%%%%%%%%
%\section{Simulation and interpretation}

\no {\bf Simulation and interpretation :}

We have simulated, as in our previous paper\cite{biasSN},
 SNIa data corresponding to the same statistical power as the data sample, where we vary the fiducial
values to study the effects of the priors (on $\Omega_M$ or/and $w_1$).

%%%\subsection{The bias effect}

We start with some illustrations of the bias introduced by the $\Omega_M$ prior 
when it is different from the
fiducial value. We consider two fiducial models which are compatible with the data, 
when no prior is applied: one in acceleration with $\Omega_M^F$ =0.5, $w_0^F$=-2.2, $w_1^F$=1.6  
and one in deceleration with $\Omega_M^F$ =0.5, $w_0^F$=-0.6, $w_1^F$=-10.
We apply the 4-fit to the two models with the two strong priors: 
$\Omega_M = 0.27 \pm 0.04$ and $\Omega_M = 0.5 \pm 0.04$.  
Figure~\ref{fig:bias} shows how the prior affects the conclusions:

\no $\bullet$ When the correct prior on $\Omega_M $ is applied, the central values are not biased but the errors 
are very large.

\no $\bullet$ When the wrong prior $\Omega_M = 0.27 \pm 0.04$ is applied, the fitted values are wrong 
but in agreement with the concordance model. The statistical errors are very small. 
In all cases, $\chi^2$ is good and does not indicate that something is wrong.

\no $\bullet$  With the data, it is not possible to distinguish between these two models, 
but the prior value can lead to wrong conclusions both on values and errors of the fitted parameters.\\

\def\figsize{5.cm}
\begin{figure} 
%\vskip-1.cm
\centerline{\epsfxsize=\figsize\epsffile{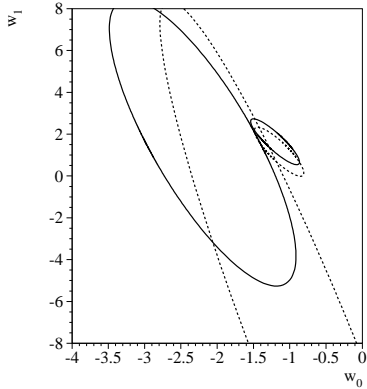}}
\vskip-1.cm
\caption[1]{\label{fig:bias}\footnotesize%
Fisher contours in the ($ w_0, w_1$) plane at $68.3\% CL$ for the two fiducial models: (in acceleration)
$\Omega_M^F$ =0.5, $w_0^F$=-2.2, $w_1^F$=1.6  and (in deceleration) $\Omega_M^F$ =0.5, $w_0^F$ =-0.6, $w_1^F$=-10.
The upper indice F is added to avoid confusion between Fiducial values and fitted values.    
The plain big and small ellipses correspond to the first model analysed  with the strong prior
$\Omega_M =0.5 \pm 0.04$ (big ellipse) or with $\Omega_M = 0.27 \pm 0.04$ as in \cite{Riess04}(small ellipse).
The dashed big and small ellipses correspond to the second model.}
%\vskip-.4cm
\end{figure}

We have then performed a complete fit analysis on the simulated data and 
scanned a large plane of fiducial values ($w_0^F,w_1^F$) with 3-fits and 4-fits,
assuming a flat universe and using two fiducial values for $\Omega_M^F$ : 0.27 or 0.5.
We always use in the fitting procedures, the strong prior $\Omega_M =0.27 \pm 0.04 $. The case 
$\Omega_M$ = $\Omega_M^F$ is equivalent to a Fisher analysis and only the errors are studied.
In the case $\Omega_M \neq \Omega_M^F$, biases are introduced in the fitted values.
\def\figsize{4.5cm}
\begin{figure}[ht]
\vspace{-.7truecm}
\begin{tabular}[t]{c c}
\centerline{\subfigureA{\epsfxsize=\figsize\epsffile{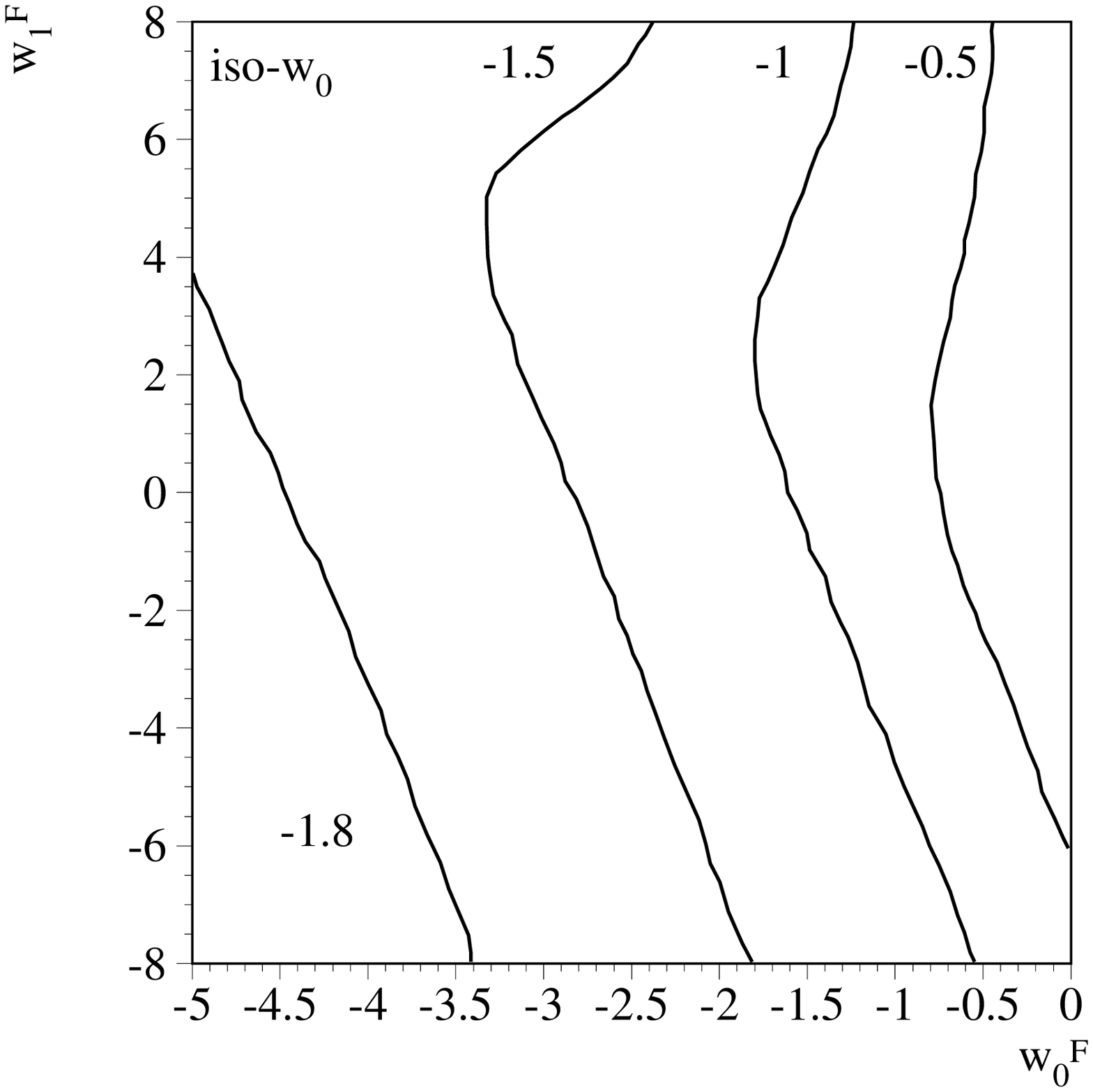}}
\subfigureA{\epsfxsize=\figsize\epsffile{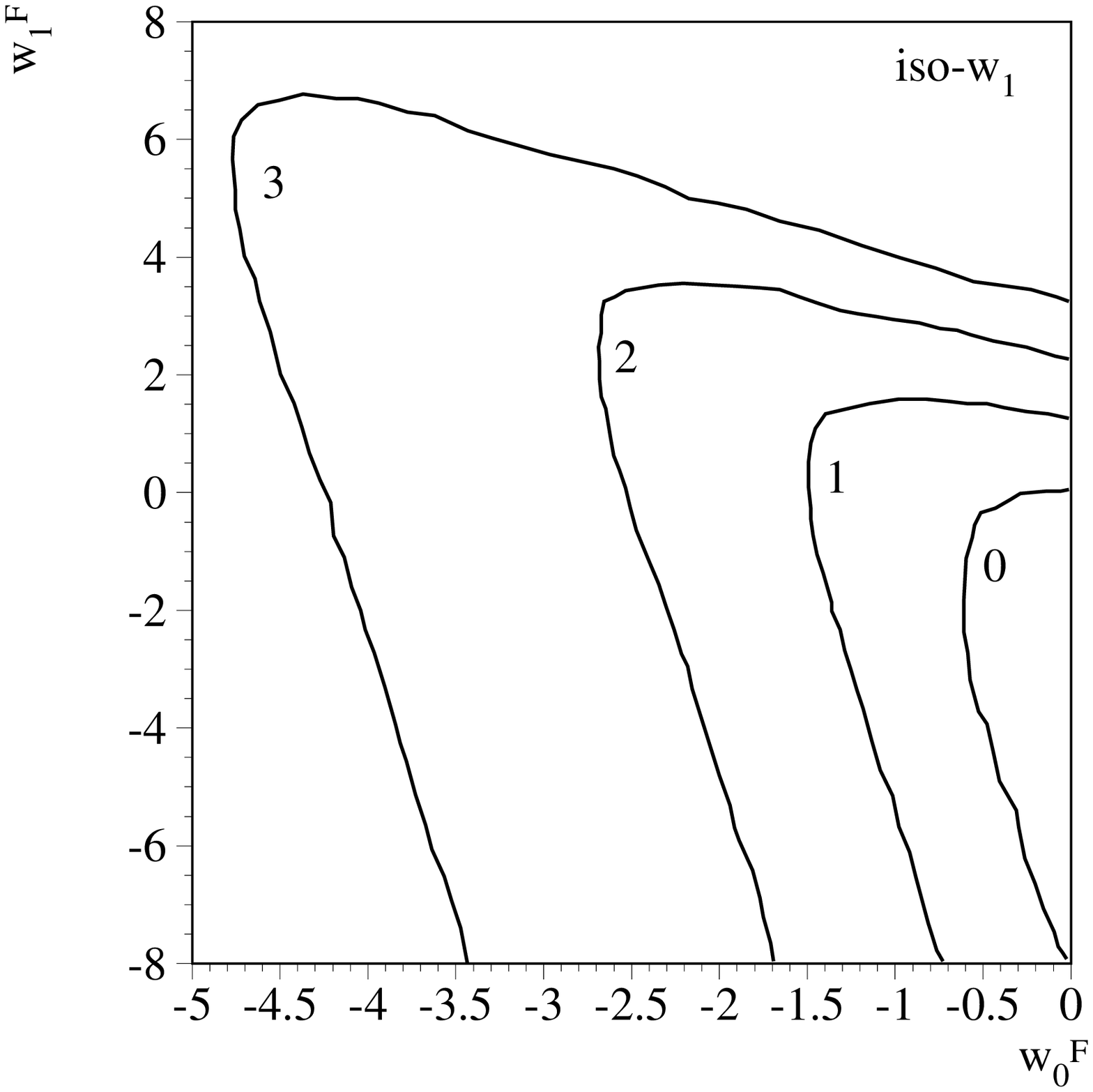}}}
\end{tabular} 
\vspace{-1.7truecm}
\caption{\footnotesize Fitted $w_0$ (left) and $w_1$ (right) iso-lines with 4-fits
for a fiducial with  $\Omega_M^F$ =0.5 and a prior $\Omega_M =0.27 \pm 0.04 $,  
in the plane ($w_0^F,w_1^F$).}
    \label{fig:fit4} 
\vspace{-.7truecm}
\end{figure}
\def\figsize{9.5cm}
 %\MTfig{fit4}{anne1.ps}{4.}{
%3 zones of fitting result for parameter $w_0$ and $w_1$ with a 4-fit procedure
%for a fiducial with  $\Omega_M$ =0.5 and a prior $\Omega_M =0.27 \pm 0.04 $,  
%in the plane ($w_0^F,w_1^F$).}

 Figure~\ref{fig:fit4} shows the fitted $w_0$ and $w_1$ iso-lines for the 4-fits
in the biased case.
The iso-lines are straight lines (not shown on the figure) when $\Omega_M^F =0.27$ (unbiased correct prior), but are
biased when $\Omega_M^F =0.5$. 
This is due to the strong correlations between $w_0$ and $\Omega_M $,
 and between $w_0$ and $w_1$. 
 
In this configuration, we observe that, for the 4-fit, when $ -5 < w_0^F < 0 $ (a relatively wide range), 
the fitted values for $w_0$ are in a narrow range centred on -1 : $-1.8<w_0<0$. For $w_1$, the situation
is even worse since with fiducial values $-8<w_1^F<8$, we get essentially positive
values for the fitted $w_1$. The actual shapes of the distortions between the fiducial and the fitted
values are readable on Fig.~\ref{fig:fit4}. 

A similar analysis performed with the 3-fit shows that the situation is even worse :
one gets $ -1.5  < w_0 < 0 $ whatever the value of $w_0^F$. As  $w_1$ is forced to $0$
and $\Omega_M$ to $0.27$,  $w_0$ is closer to  -1 
which corresponds to the preferred solution for the fit. 

One can illustrate further this very problematic point,
by defining ``confusion contours'', namely some contours 
which identify the models in the fiducial parameter space ({\it e.g.} ($w_0^F,w_1^F$))
that could be confused with another model. For instance, the contours of Figure~\ref{fig:conf}
give the models in the plane ($w_0^F,w_1^F$) with $\Omega_M^F=0.5$ that can be 
confused (at 1 and 2$\sigma$) with the
concordance model if the (wrong) strong prior is applied. The two models used 
for the illustrative Fig.~\ref{fig:bias}
are taken from extreme positions in this confusion contour of Fig.~\ref{fig:conf}.

For the 3-fit, the confusion contours with the concordance model
are very large and include all models having roughly $w_1^F < (-5w_0^F-10)$. 
The situation here is particularly bad since the fitting procedure is making
two strong assumptions ($w_1=0$ and $\Omega_M =0.27\pm 0.04$) which are not verified by
the fiducial cosmology (two biases).

\def\figsize{4.5cm}
\begin{figure}[ht]
\vspace{-.7truecm}
\begin{tabular}[t]{c c}
\centerline{\subfigureA{\epsfxsize=\figsize\epsffile{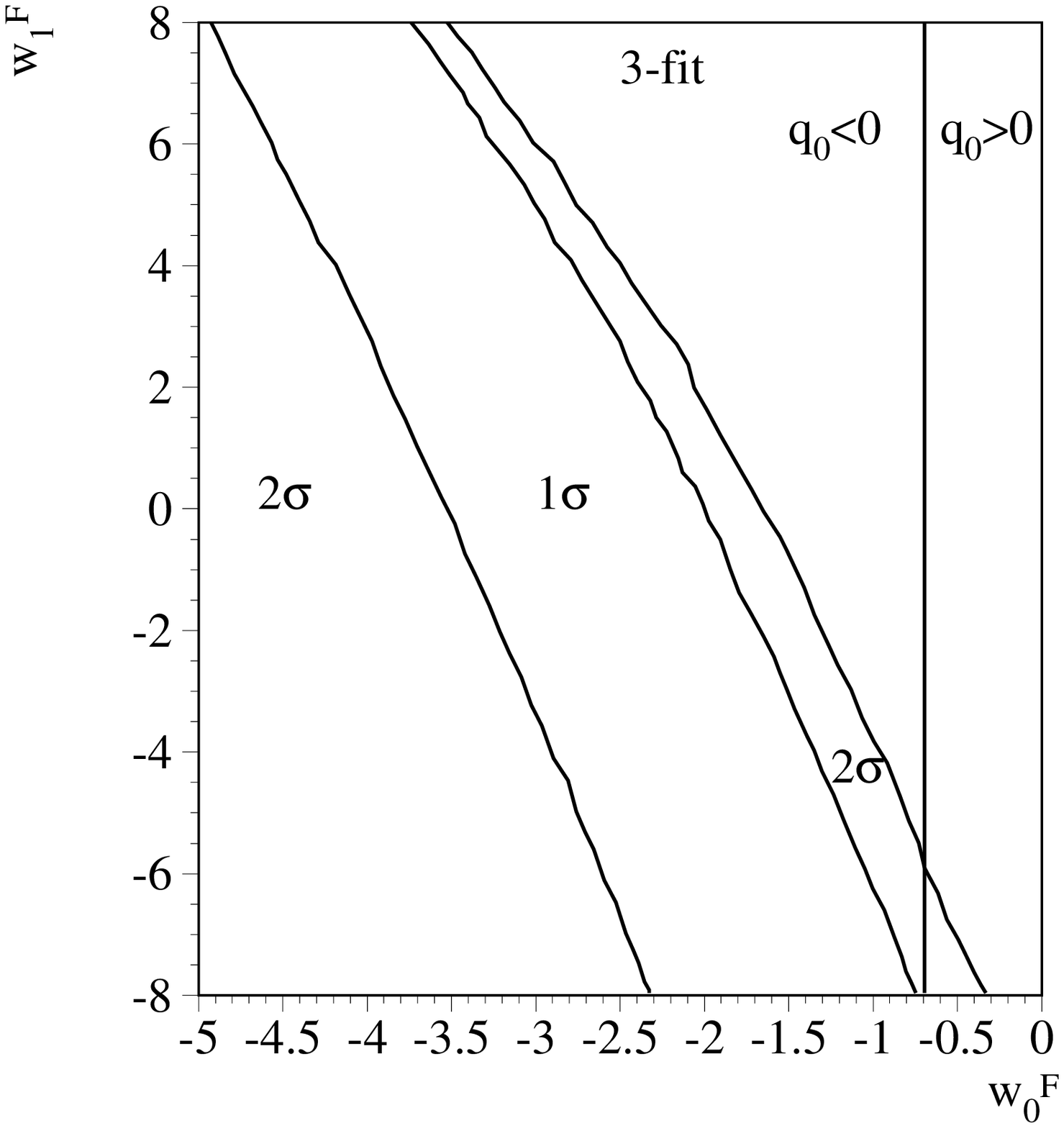}}
\subfigureA{\epsfxsize=\figsize\epsffile{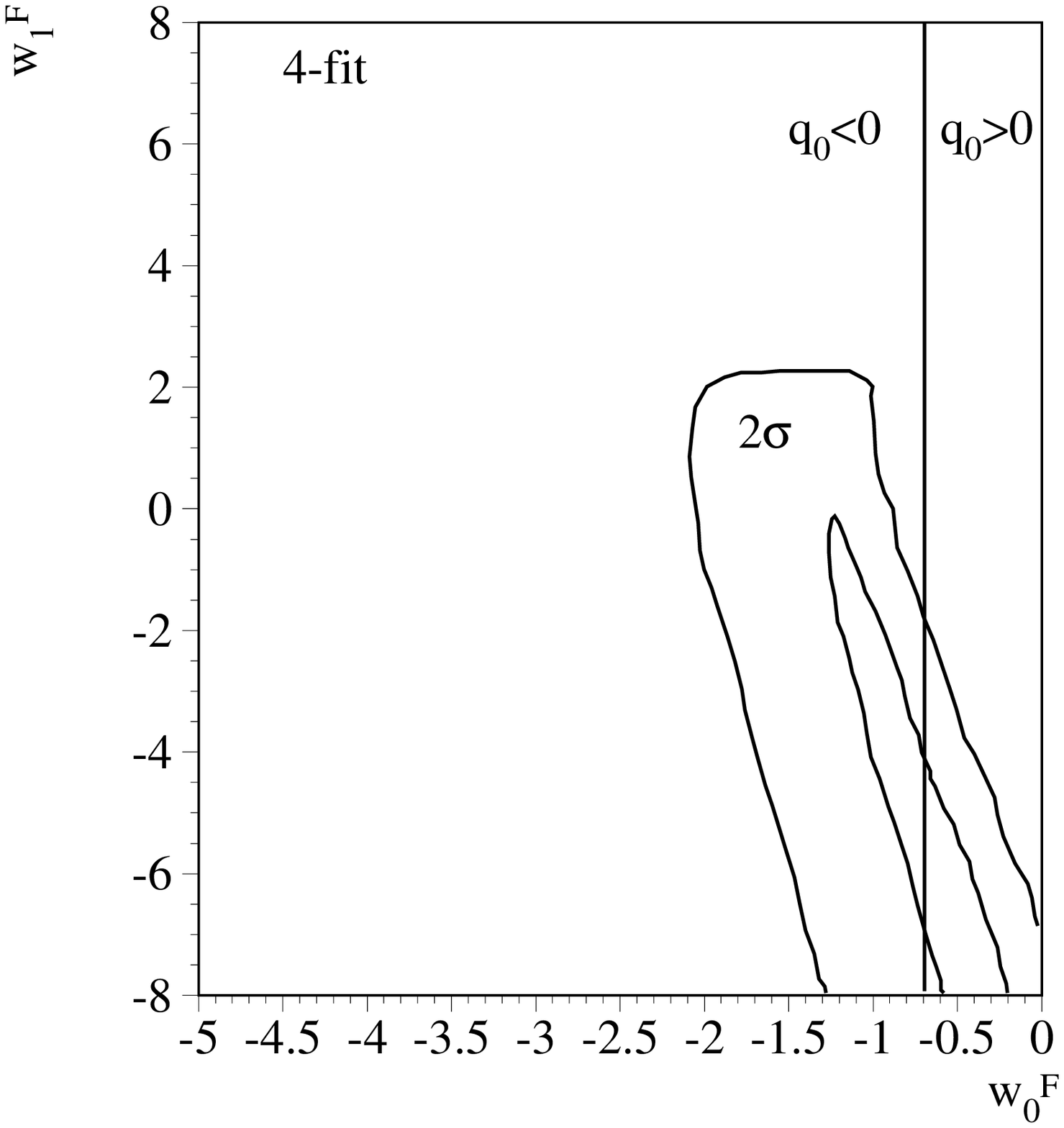}}}
\end{tabular} 
\vspace{-1.7truecm}
\caption{\footnotesize Confusion contours in the fiducial plane ($w_0^F,w_1^F$) which identify the models
that would be confused with the concordance model at 1 and 2$\sigma$ for the 3-fit(left) and the 4-fit(right)
procedures. The strong prior is $\Omega_M =0.27 \pm 0.04 $ whereas the fiducial model is $\Omega_M^F = 0.5$.
The vertical line separates accelerating from decelerating models.}
    \label{fig:conf} 
\vspace{-.2truecm}
\end{figure}
\def\figsize{9.5cm}
%

%The fit is thus biased as soon as one of the values used inside the procedure 
% is wrong. The simulation can reproduce results in agreement with the present data based on models with 
%$\Omega_M^F $=0.5 and almost all possible values in the ($w_0^F $, $w_1^F$) plane. The concordant model
%is a solution for a large fraction of these models.

%%% \subsection{Parameter error analysis}

The next step is to study the parameter errors. We look at the correlation of the 
errors using fiducial models where $\Omega_M^F$ =0.27 
or 0.5.
We  determine the $w_0$ and $w_1$ errors, scanning the full plane ($w_0^F,w_1^F$) 
using 4-fits.

Some regions of the parameter space (see Figure~\ref{fig:fisher}) are favoured and always produce 
small errors. This is due to the correlation between $w_0$ and $w_1$. 
The error depends strongly on the fitted $w_0$ and $w_1$ 
values but not strongly  on the $\Omega_M $ value: a different value of $\Omega_M $ affects
the scale of the errors but not the shape of the plots. We find a linear scaling of the error when 
we change $\Omega_M $ from 0.27 to 0.5 ({\it i.e.} a factor 2). 

\def\figsize{4.5cm}
\begin{figure}[ht]
\vspace{-.7truecm}
\begin{tabular}[t]{c c}
\centerline{\subfigureA{\epsfxsize=\figsize\epsffile{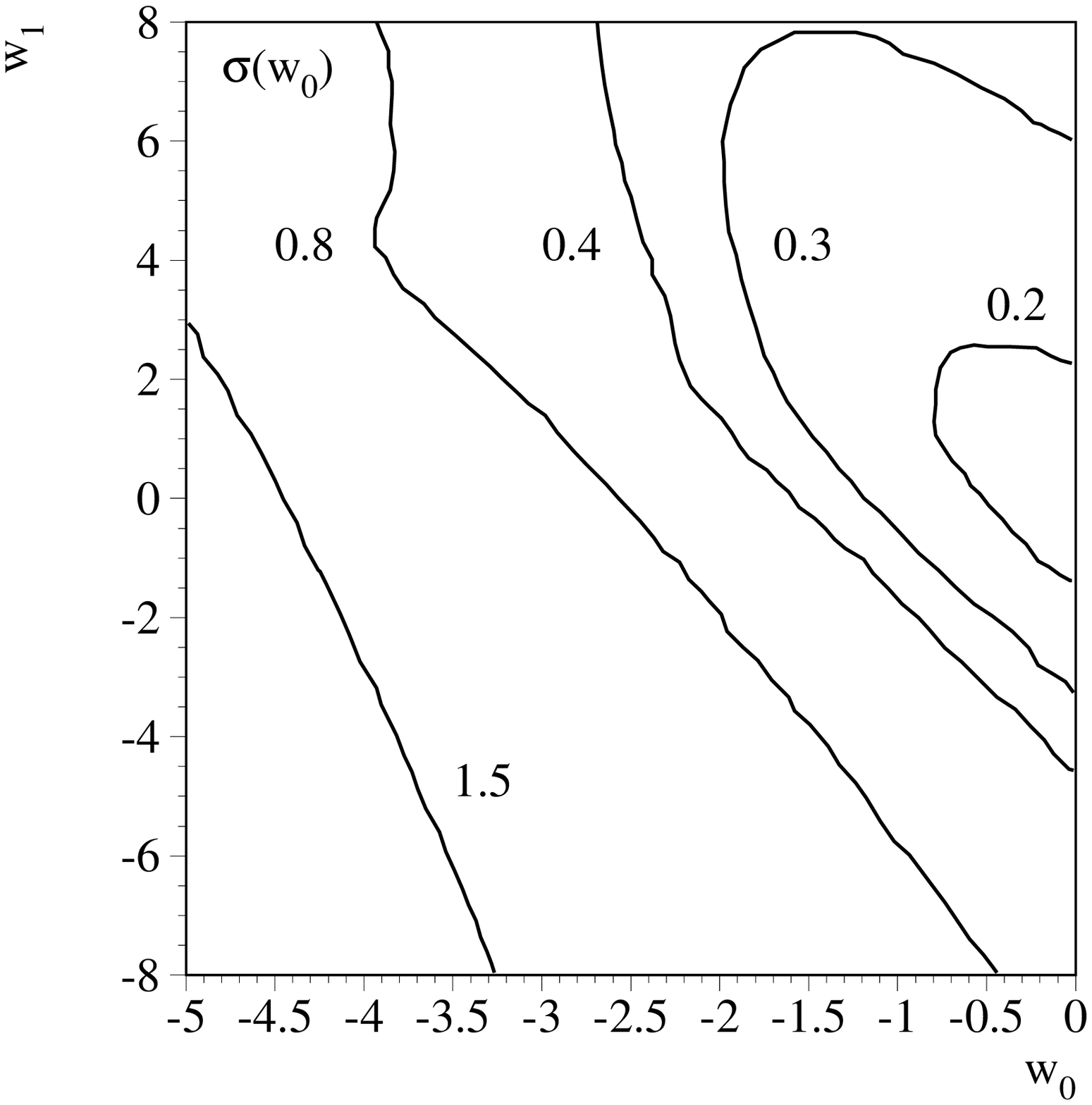}}
\subfigureA{\epsfxsize=\figsize\epsffile{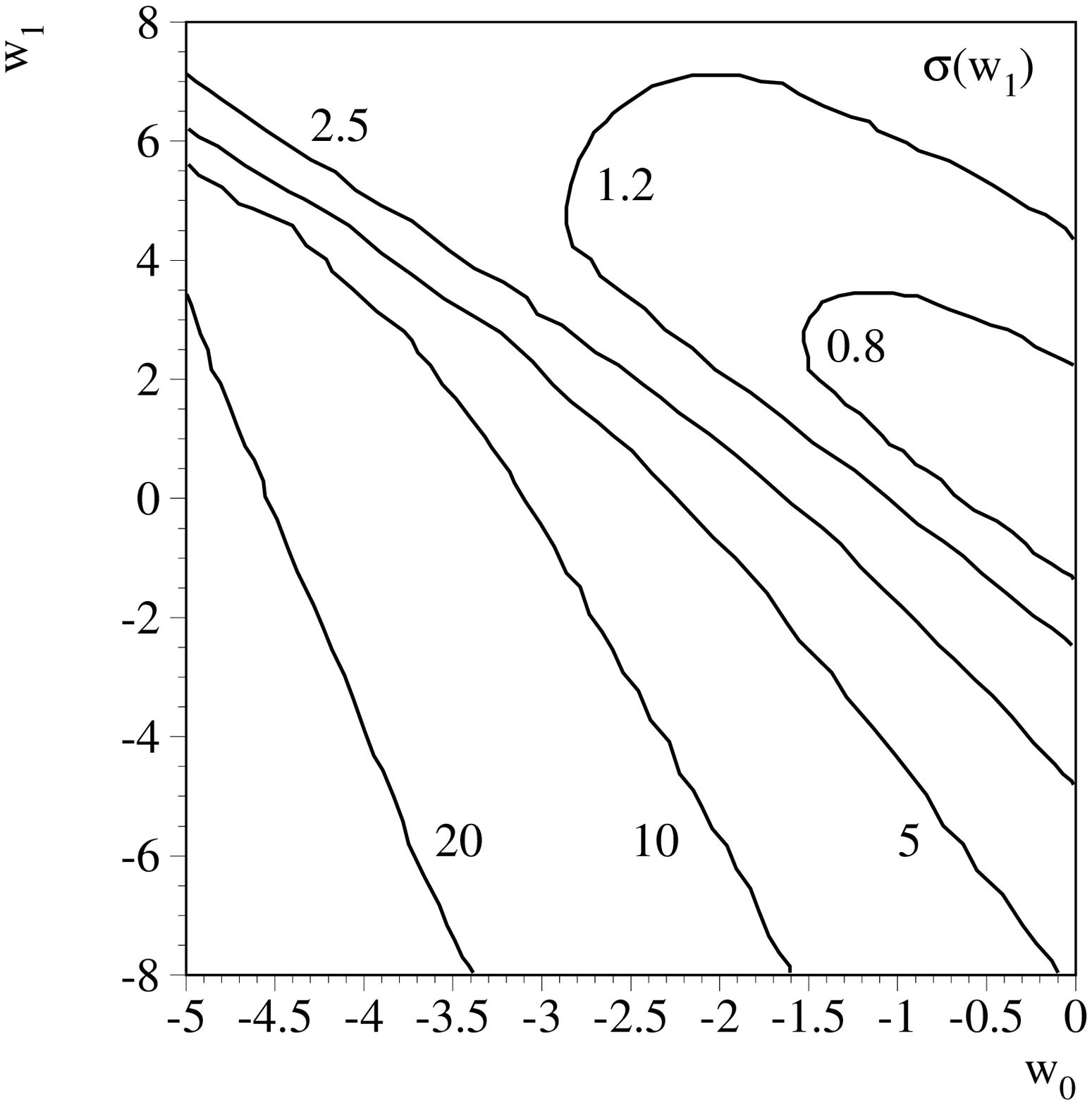}}}
\end{tabular} 
\vspace{-1.7truecm}
\caption{\footnotesize $w_0$ and $w_1$ errors for a 4-fit
with the correct strong prior, for a fiducial with $\Omega_M^F$ =0.27.}
    \label{fig:fisher} 
\vspace{-.2truecm}
\end{figure}
\def\figsize{9.5cm}

Combining this with the previous paragraph leads to an interesting point, {\it i.e.} 
the favoured fitted values of the fit ($w_0> -1.8$ and $w_1> 0$), which were shown to be 
mainly driven by the prior value, correspond also to the region of the plane 
where the parameters errors are always small.
%For these values, the results always have $\sigma(w_0) < 0.8$ and $\sigma(w_1) < 1.5$ 
%whatever the value of $\Omega_M^F$. 
  
We conclude that the applied fitting procedure with this strong prior can bias
the conclusions by constraining the ($w_0,w_1$) solution near the (-1,0) 
solution, where the statistical 
error is always very small. 
In particular, Riess et al.\cite{Riess04} found a gain factor of order
8 on the accuracy of the neasurement compared to previous analysis.
This is mainly due to the $\Omega_M$ prior and not to the inclusion of the high $z$
HST events.
The present observational constraints on $\Omega_M$ are thus an important issue. \\

\no {\bf Revisited conclusions on existing data :}

Most reviewers of cosmological parameters favour a value close to the strong prior 
choice made by \cite{Riess04}. This result is based on WMAP\cite{WMAP} data 
combined with 2dF data\cite{2dF} or more recently with SDSS data\cite{SDSSWMAP},
and corresponds to \cite{WMAP}(\cite{SDSSWMAP}) $\Omega_M$ = 0.27(0.3) $\pm$ 0.04
with $h= 0.71(0.70)^{+0.04}_{-0.03}$.
%However, Spergel et al.\cite{WMAP} mention that a solution with  $\Omega_M$ = 0.47, $w =-0.5$ and
%$h =0.57$ in the CMB, "has a nearly identical angular power spectrum as the $\Lambda CDM$ model".
However, these results are based on several prior assumptions in order to lift the degeneracies among
the various cosmological parameters ({\it e.g. } $\Omega_M$, $h$, $\sigma_8$, $w$ ...).
For instance, Spergel et al.\cite{WMAP} mention that a solution with  $\Omega_M$ = 0.47, $w =-0.5$ and
$h =0.57$ in the CMB is degenerate with the $\Lambda CDM$ model.
This kind of
solution is excluded for three reasons: the Hubble Constant value is 2$\sigma$ lower 
than the HST Key Project value and the model is a poor fit to the 2dF and SN data. 

However, {\it i)} in spite of the precise HST result, the Hubble constant value is still 
controversial\cite{Blanchard, Conversi}. 
{\it ii)} We have shown in the previous section that the SN data 
analysis can only 
conclude that $\Omega_M < 0.6$ (see Fig.~\ref{fig:omdata}).
{\it iii)} The 2dF and the SDSS Collaborations \cite{2dF,SDSSPk} have extracted $\Omega_M h$ from an analysis
of the power spectrum of galaxy redshift surveys. The degeneracy between $\Omega_M h$ and the baryon
fraction is lifted thanks mainly to the precise determination of the baryon fraction
by CMB data (see Fig.38 of \cite{SDSSPk}). Should that prior change, the preferred values
from SDSS would indicate a higher value of $\Omega_M h$.

In addition,
a large variety of observations give constraints on $\Omega_M$, which is found to vary
from 0.16 $\pm$ 0.05 \cite{BahcallML}
up to a value above 0.85 \cite{Vauclair}.

Conversi et al.\cite{Conversi} provide an interesting critical analysis on the present
constraints on cosmological parameters, especially on $\Omega_M$, $h$, and $w$. Through the study of the
degeneracies, they show that the result  $\Omega_M = 0.27\pm 0.04$ is obtained under
the assumption of the $\Lambda CDM$ model, and provide specific examples with smaller 
$h$ ($h<0.65$) and higher $\Omega_M$ 
($\Omega_M>0.35$) which are in perfect agreement with the most recent CMB and galaxy redshift surveys.

In conclusion, we follow the point of view of Bridle et al.\cite{Bridle}, 
who argue that it may be "that the real uncertainty is much greater" than
the 0.04 error obtained from the combination of CMB and large scale structure data.\\

Returning to SN data analysis,
we suggest, for the time being,
to reevaluate the conclusions by relaxing the $\Omega_M$ central value.
Figure~\ref{fig:result} shows the $95\% CL$ constraints in the ($w_0$,$w_1$) plane obtained
from the gold sample \cite{Riess04} with no prior assumption on $\Omega_M$. 
%(the contour with the strong prior of \cite{Riess04} is also shown).
Taking an uncertainty of 0.2, which is the intrinsic sensitivity of SN results (see Table I, line 6),
does not change the conclusions:

\def\figsize{7.cm}
\begin{figure} 
\vskip-1.5cm
\centerline{\epsfxsize=\figsize\epsffile{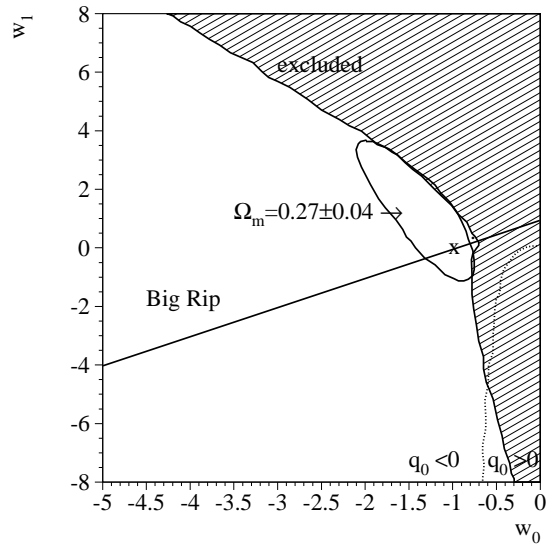}}
\vskip-1.5cm
\caption[1]{\label{fig:result}\footnotesize%
The shaded region is excluded at $95\% CL$ when no prior on $\Omega_M$ is applied.
The ellipse corresponds to the strong prior constraints as in \cite{Riess04}.
The (x) is $\Lambda CDM$. The dotted line separates accelerating from decelerating models.
}
%Models with $w_1>w_0+1$ will end in a Big Rip.}
\vskip-0.6cm
\end{figure}

\no $\bullet$ 
Large positive variations in time of the equation of state are excluded (at $95\% CL$)
since the dark energy density blows up as $e^{3w_1z}$ \cite{WangTegmark}.

\no $\bullet$ The quintessence models which have in general  ($w_0>-1$, $1>w_1>0$) \cite{WA} 
are seriously constrained.
%The quadrant ($w_0>-1$, $w_1>0$) is seriously constrained and it corresponds to quintessence
%models \cite{WA} (which have in general $w_1<1$). 
For instance, the SUGRA model \cite{SUGRA}
characterized by $w_0 \approx -0.8$ and $w_1\approx 0.3$\cite{WA} is close to the border of the $95\% CL$ contour
(precisely, one gets $\Delta {\chi}^2$=3.5 corresponding to an exclusion at $80\% CL$). 
%This zone corresponds to large fitted $\Omega_M$ values. 

\no $\bullet$ The quadrant ($w_0>-1$, $w_1<0$) corresponding to k-essence models \cite{kessence} or
some Big Crunch models \cite{Kallosh,Riess04}, is not the ``least favoured'', contrary to the
conclusions drawn with the strong prior \cite{Riess04}. We find that if $w_0$ goes
towards 0, then $w_1$ should be more and more negative.

\no $\bullet$ If $w_0<-1$, the constraints on $w_1$ are weak (except for large positive values). This region
of the plane corresponds to phantom models \cite{Caldwell} which have unusual properties and may have very
different consequences for the fate of the Universe 
({\it e.g.} models with $w_1>w_0+1$ will end in a Big Rip \cite{WangTegmark}).
Models with very exotic $w(z)$ may come from modified gravity\cite{LinderG}.
The class of models with $w_1<0$ is roughly excluded
at $95\% CL$, if the strong prior $\Omega_M=0.27\pm0.04$ is used \cite{Riess04}, but is perfectly allowed
for higher $\Omega_M$ values (or larger prior errors).

\no $\bullet$  As can be seen on Fig. 6 (and also on Figures 1, 4 and
the decelerating model used to draw
Fig.2), our analysis without assumptions on $\Omega_M$ and $w_1$, allows decelerating models
with specific properties : low $w_0$, $\Omega_M\approx 0.5$ and $w_1<<0$.
One can wonder if this result is not in tension with the geometrical test performed in
\cite{Riess04} where the only assumption is to use a linear functional form for $q(z)$ ({\it i.e.} $q_0+q_1z$).
It can be shown that a varying equation of state implies a non-linear
$q(z)$, in particular, the linear approximation breaks down if $w_1^F<-1.5$.
More details on this more subtle analysis will be presented in a forthcoming paper \cite{newrenoir}.\\

To go further, a coherent combined analysis of all data is mandatory,
with a proper treatment of correlations and no prior assumptions.
Some recent papers go in that direction \cite{WangTegmark,Melchiorri,Hannes,Corasaniti}.

In addition, as soon as the statistical errors will become smaller, systematic questions cannot be neglected
 and should be controlled at the same level of precision. 
This is the  challenge for the next generation of experiments. 
A promising approach is to combine SNIa with weak lensing,
as proposed by the future dedicated SNAP/JDEM mission \cite{LinderSNAP}.\\

\no {\bf Acknowledgments :} 

We thank the members of the cosmology group of the Laboratoire d'Astrophysique de Marseille (LAM),
and in particular C. Marinoni for helpful discussions.
$^*$``Centre de Physique Th\'eorique'' is UMR 6207 - ``Unit\'e Mixte
de Recherche'' of CNRS and of the Universities ``de Provence'',
``de la M\'editerran\'ee'' and ``du Sud Toulon-Var''- Laboratory
affiliated to FRUMAM (FR 2291).
$^+$``Centre de Physique des Particules de Marseille'' is UMR 6550 
of CNRS/IN2P3 and of the University
``de la M\'editerran\'ee''.
%%%%%%%%%%%

%%%%%%
%\newpage
%\bibliographystyle{unsrt}

\end{document}